\documentclass[onefignum, onetabnum,onealgnum,onethmnum, oneeqnum]{siamonline190516}
\usepackage{comment}
\usepackage{lipsum}
\usepackage[margin=1in, footskip=36pt]{geometry}

\usepackage{endnotes}

\usepackage{titlesec}
\usepackage{titletoc}
\usepackage{pgfplotstable}
\titleclass{\task}{straight}[\section]
\newcounter{task}
\renewcommand{\thetask}{\arabic{task}}
\titleformat{\task}[hang]
    {\normalfont\LARGE\bfseries}{Task \thetask:}{1em}{}

\titleformat*{\task}{\color{header1}\bfseries}

\titlecontents{task}
              [3.8em] 
              {}
              {\contentslabel{2.3em}}
              {\hspace*{-2.3em}}
              {\titlerule*[1pc]{.}\contentspage}

\titlespacing*{\section}{0ex}{1ex}{1ex}
\titlespacing*{\subsection}{0ex}{1ex}{1ex}
\titlespacing*{\subsubsection}{0ex}{1ex}{1ex}
\titlespacing*{\paragraph}{0ex}{1ex}{1ex}
\titlespacing*{\subparagraph}{0pt}{1ex}{1ex}
\titlespacing*{\task}{0em}{1ex}{1ex}

\usepackage[sort&compress,comma,square,numbers]{natbib}

\usepackage{amsfonts,amsmath,amssymb}
\usepackage{bbm}

\usepackage{multicol}
\usepackage{enumitem}
\setlist[enumerate]{wide, labelindent=1cm,  noitemsep}
\setlist[itemize]{noitemsep}
\setlist[description]{noitemsep}

\usepackage{hhline}

\usepackage{todonotes}
\RequirePackage{ulem}

\definecolor{tableheadcolor}{gray}{0.92}
\newcommand{\topline}{ %
        \arrayrulecolor{maroon}\specialrule{0.1em}{\abovetopsep}{0pt}%
        \arrayrulecolor{tableheadcolor}\specialrule{\belowrulesep}{0pt}{0pt}%
        \arrayrulecolor{maroon}}
\newcommand{\midtopline}{ %
        \arrayrulecolor{tableheadcolor}\specialrule{\aboverulesep}{0pt}{0pt}%
        \arrayrulecolor{maroon}\specialrule{\lightrulewidth}{0pt}{0pt}%
        \arrayrulecolor{white}\specialrule{\belowrulesep}{0pt}{0pt}%
        \arrayrulecolor{maroon}}
\newcommand{\bottomline}{ %
        \arrayrulecolor{white}\specialrule{\aboverulesep}{0pt}{0pt}%
        \arrayrulecolor{maroon} %
        \specialrule{\heavyrulewidth}{0pt}{\belowbottomsep}}%

\pgfplotstableset{normal/.style ={%
        header=true,
        string type,
        font=\small,
    	column type={p{.092\textwidth}},
        every head row/.style={
            before row={\topline\rowcolor{tableheadcolor}},
            after row={\midtopline}
        },
        every odd row/.style={
            before row={\rowcolor{maroon}}
        },
        every last row/.style={
            after row=\bottomline
        },
        col sep=&,
        row sep=\midrule
    }
}

\usepackage{multirow}

\usepackage{booktabs}
\usepackage{fancyhdr}
\usepackage{fullpage}

\RequirePackage{titlesec}
\RequirePackage[font={footnotesize},{singlelinecheck=false}]{caption}
\usepackage[T1]{fontenc}
\usepackage[scaled]{helvet}
\usepackage[scaled=1.10, med]{zlmtt}

\usepackage[ruled,linesnumbered]{algorithm2e}

\usepackage{caption}
\usepackage{placeins}

\usepackage[symbol]{footmisc}

\usepackage{ericmath}
\usepackage[createShortEnv]{proofs}
\usepackage{xcolor}

\usepackage{tcolorbox}
\usepackage{thmtools}
\usepackage{etoolbox}  

\newcounter{rulenum}
\newcommand{\airule}[1]{%
  \refstepcounter{rulenum}%
  \par%
  \vspace{0.2ex}%
  \noindent%
  {\bfseries\color{header1} Rule \arabic{rulenum}: #1}%
}

\makeatother

\usepackage{mathrsfs}
\usepackage[scr=rsfs,cal=boondox]
{mathalfa}

\title{Ten Simple Rules for AI-Assisted Coding in Science}

\author{
    Eric W.~Bridgeford$^{1,\dagger}$,
    Iain Campbell$^2$,
    Zijiao Chen$^1$,
    Zhicheng Lin$^{3,4}$,
    Harrison Ritz$^2$,
    Joachim Vandekerckhove$^5$,
    Russell A. Poldrack$^{1}$.
    \thanks{
     $^1$ Stanford University, Stanford, CA, USA,
     $^2$ Princeton University, Princeton, NJ, USA,
     $^3$ University of Science and Technology of China, Hefei, China,
     $^4$ Yonsei University, Seoul, Republic of Korea,
     $^5$ University of California, Irvine, CA, USA \\
    $^\dagger$ Corresponding author:
      Eric W.~Bridgeford (\email{ericwb@stanford.edu}).}
}

\makeatletter
\def\thanks#1{\protected@xdef\@thanks{\@thanks
        \protect\footnotetext{#1}}}
\makeatother
\makeatletter
\renewcommand\@biblabel[1]{#1.}
\makeatother

\usepackage{lineno}
\begin{document}

\maketitle
\begin{abstract}
While AI coding tools have demonstrated potential to accelerate software development, their use in scientific computing raises critical questions about code quality and scientific validity. In this paper, we provide ten practical rules for AI-assisted coding that balance leveraging capabilities of AI with maintaining scientific and methodological rigor. We address how AI can be leveraged strategically throughout the development cycle with four key themes: problem preparation and understanding, managing context and interaction, testing and validation, and code quality assurance and iterative improvement. These principles serve to emphasize maintaining human agency in coding decisions, preserving the domain expertise essential for methodologically sound research. These rules are intended to help researchers harness AI's transformative potential for faster software development while ensuring that their code meets the standards of reliability, reproducibility, and scientific validity that research integrity demands.
\end{abstract}
\section{Introduction}

The integration of artificial intelligence into scientific computing represents one of the most significant shifts in research methodology since the advent of personal computers. Large language models (LLMs) trained on vast corpora of code can now generate syntactically correct, functionally appropriate programs from natural language descriptions, a capability that was inconceivable just a few years ago \cite{chen2021evaluating}. Tools like GitHub Copilot, ChatGPT, and Claude have democratized access to sophisticated programming assistance, enabling researchers with limited coding experience to implement complex analyses and build robust scientific software \cite{peng2023impact}. Agentic coding tools like Claude Code and Cursor have further enabled entire coding workflows by invoking tools outside the language model.

AI-assisted coding tools have demonstrated measurable productivity gains in some controlled studies, with benefits spanning development speed, code quality, and maintainability \cite{peng2023impact}. However, the evidence for these benefits remains contested and situation-dependent. While some enterprise studies and developer surveys report significant productivity increases and improved code quality \cite{kalliamvakou2024measuring}, recent randomized controlled trials with experienced developers found that AI tools actually slowed completion times, despite developers believing they were working faster \cite{becker2025measuring}. Additional concerns about code quality have emerged, with research analyzing almost 200 million lines of code showing substantial increases in copy-pasted code and decreases in refactoring as AI use has become more prominent when using AI assistants \cite{gitclear2024}. These contradictory findings suggest that productivity effects are far from well-understood, and may vary based on developer experience, task complexity, and codebase characteristics. These questions become particularly acute in scientific computing, where code is not merely a means to an end but often embodies scientific reasoning, methodological decisions, and domain expertise. The validity, reproducibility, and interpretability of scientific software directly impact research integrity and the reliability of scientific findings \cite{poldrack2024better}.

The implications for scientific computing are profound. Programming involves complex problem decomposition, algorithmic thinking, and domain-specific reasoning. These cognitive skills that atrophy with excessive AI dependence. Furthermore, scientific code often requires deep understanding of mathematical models, statistical methods, and domain-specific conventions that cannot be adequately captured by AI tools trained on general programming corpora. These challenges are compounded by the technical limitations inherent to current AI systems: their context windows constrain how much code they can process simultaneously, their stateless nature means they forget previous interactions, and their tendency toward ``context rot'' \cite{chroma2024context} can cause them to lose track of important details even within their processing limits.

Effective use of AI coding tools requires understanding how to work within and around these constraints. Techniques like strategic prompting, test-driven development, and externally-managed context files (such as memory files and constitution files) can help maintain consistency across AI interactions. Different tools (from conversational interfaces to interactive coding assistants to autonomous coding agents) each offer distinct capabilities and limitations that must be matched to specific development tasks. (For readers unfamiliar with these concepts, we provide detailed definitions in Appendix \ref{app:background}.)

The rules presented in this paper emerge from our collective experience using AI-assisted coding tools, and highlight both the substantial promise and documented risks of AI-assisted coding in scientific settings. We hope they provide a framework for harnessing AI's transformative potential while preserving the methodological rigor and domain expertise essential for high-quality scientific computing. These guidelines emphasize the importance of maintaining human agency in the coding process, establishing robust testing and validation procedures, and strategically managing the interaction between human expertise and AI assistance.

\paragraph{Who is this paper for?} These guidelines are intended for anyone who develops scientific software that will be used more than once, whether by themselves, their collaborators, or the broader research community. This includes both scientists who primarily use code to generate research outputs and developers who build reusable tools and packages. Our focus is on creating maintainable, reliable software rather than one-off scripts. If you write code that needs to work reliably and repeatably, be understood by others, or be built upon in the future, we believe these rules are for you.

\section{The rules} 

To illustrate these rules with concrete examples, we provide an interactive Jupyter Book demonstrating both effective and ineffective implementations of each rule (3+ examples per rule), available at \href{https://poldracklab.org/10sr_ai_assisted_coding}{poldracklab.org/10sr\_ai\_assisted\_coding} and permanently indexed at \cite{bridgeford2025aiassist}. These worked examples with detailed analysis will reinforce readers' understanding of what good and flawed AI interactions look like, and put the strategies discussed herein into practice.

\subsection{Preparation and Understanding} The effectiveness of AI-assisted coding is fundamentally constrained by the clarity and completeness of what you bring to the interaction. AI models cannot compensate for gaps in your understanding of the problem domain, your inability to recognize appropriate solutions, or your unfamiliarity with relevant tools and conventions. These limitations arise because AI tools lack true domain expertise. They pattern-match from training data rather than reason from first principles, and they cannot assess whether their outputs align with field-specific best practices you haven't mentioned. Without this foundational preparation, you risk ``vibe coding,'' accepting AI-generated code you cannot evaluate, debug, or maintain.

\airule{Gather Domain Knowledge Before Implementation} \label{rule:gather} Know your problem space before writing code. Understand data shapes, missing data patterns, field-specific libraries, and existing implementations that could serve as models. You don't need to be an expert initially; use AI to help research domain standards, available datasets, common approaches, and implementation patterns before diving into coding. This reconnaissance phase prevents you from reinventing wheels or violating field conventions. Share your current understanding level with the AI and iteratively build context through targeted questions about tools, data structures, and best practices, asking for specific references and paper summaries. This upfront investment ensures that your code aligns with community standards and handles real-world data appropriately.

\airule{Distinguish Problem Framing from Coding} \label{rule:framing_vs_coding} Framing a problem in a programmatic way and coding are not the same thing \cite{martin2008clean}. Programmatic problem framing is problem solving: understanding the domain, decomposing complex problems, finding the right levels of abstraction, designing algorithms, and making architectural decisions. Coding is the mechanical translation of these concepts into executable syntax in a programming language. Using AI coding tools effectively requires that you deeply understand the problem from a programmatic perspective that you are trying to solve; in most cases this understanding transcends the particular programming language, and the actual code implementation itself. AI tools excel at coding tasks, generating syntactically correct implementations from well-specified requirements, but they currently require human guidance for programmatic problem framing decisions that involve domain expertise, methodological choices, and scientific reasoning. You can't effectively guide or review what you don't understand, so establish fluency in at least one programming language and fundamental concepts before leveraging AI assistance. This foundation allows you to spot when generated code deviates from best practices or introduces subtle bugs. Without this knowledge, you're essentially flying blind, unable to distinguish between elegant solutions and convoluted workarounds.

\airule{Choose Appropriate AI Interaction Models} \label{rule:choice_of_model} It's tempting to use the AI tools to independently generate a complete codebase, but one quickly ends up being separated from the code and making mistakes. A pair programming model, where one directs interactive AI assistants through comments in the code, can be a way to stay in close touch with the evolution of the codebase. The utility of different language models and tool types will depend on the specific development tasks, developer preferences, and project constraints. A summary of different interaction paradigms, as well as their strengths and limitations in 2025, are provided in Table \ref{table:paradigms}.

\begin{table}[h]
\centering
\begin{tabular}{|l|p{4.5cm}|p{7cm}|}
\hline
\textbf{Tool Type} & \textbf{Best For} & \textbf{Description} \\
\hline
\parbox[t]{3.5cm}{\textbf{Conversational}\\(ChatGPT, Claude)} & Architecture design, complex debugging, learning new concepts & Deep reasoning and flexible problem-solving with extensive context handling, but requires manual code transfer and loses context between sessions \\
\hline
\parbox[t]{3.5cm}{\textbf{IDE Assistant}\\(Copilot, IntelliSense)} & Code completion, refactoring, maintaining flow & Seamless workflow integration with immediate feedback and preserved code context, but limited reasoning for complex architectural decisions \\
\hline
\parbox[t]{3.5cm}{\textbf{Autonomous Agent}\\(Cursor, Claude Code, Aider)} & Rapid prototyping, multi-file changes, large refactoring & High-speed implementation that can work independently across multiple files, but risks code divergence and requires careful monitoring \\
\hline
\end{tabular}
\caption{AI-assisted development tools are categorized by interaction model and deployment scenario. Each paradigm offers distinct advantages for different phases of software development, with trade-offs between automation level and developer control.}
\label{table:paradigms}
\end{table}

\subsection{Context Engineering \& Interaction} Once you understand your problem domain and have chosen appropriate tools, effective AI collaboration requires careful management of how you structure prompts and maintain context\footnote[4]{To avoid confusion, we use the word “context” throughout this article only when referring to information that is relevant to the AI system's operation or your interaction with it. This includes concepts like in-context learning, context management, and context windows, or any general information that is directly relevant to the problem you are trying to solve using AI. Where the word might have otherwise appeared in more general usages (e.g., ``in the context of'', ``scientific contexts'', etc.) we have used alternate terms for clarity. } across interactions. Most AI systems are stateless, forgetting previous conversations, while others struggle with context rot as conversations grow long. This creates two critical challenges: ensuring the AI has all necessary information to generate appropriate code, and recognizing when a conversation has become too polluted with failed attempts to be productive. 

\airule{Start by Thinking Through a Potential Solution}\label{rule:thinking} Begin AI coding sessions by first working to understand and articulate the problem you're trying to solve, specified at a level of abstraction that makes it solvable by code, and think through how you anticipate it might be solved. Think through the entire problem space: What are the inputs and expected outputs? What are the key constraints and edge cases? What does success look like? This problem-focused approach serves a dual purpose: it helps you clarify exactly what you want the AI to accomplish so that you can evaluate its outputs appropriately, and it prevents the AI from making incorrect assumptions about your goals. When you provide problem context along with architectural details of how you anticipate a solution working (i.e., how it might fit in with the ``bigger picture’’), the AI generates code that fits naturally into your project rather than creating isolated solutions. Include details about data flow, component interactions, and expected interfaces when possible or if relevant. This approach transforms the AI from a code generator into an architecture-aware development partner. You can use LLMs to help generate externally-managed context files, and also look at GitHub Spec Kit for specification-driven workflows that define project requirements and gated phases (Specify, Plan, Tasks). AI can help you implement sophisticated patterns like structured checklists for iterative development that would be tedious to write from scratch.

\airule{Manage Context Strategically} \label{rule:context} Context (which for LLMs refers to all of the information currently in the model’s equivalent of ``working memory’’) is everything in AI-assisted coding. Provide all necessary information upfront through clear documentation, attached references, or structured project files with dependencies included. Don't assume the AI retains perfect context across long conversations; explicitly restate critical requirements, constraints, and dependencies when interactions get complex. Keep track of context and clear or compact when it's getting close to limits. Use externally-managed context files to keep important context available across sessions while minimizing irrelevant details that can degrade AI performance. Agents can effectively use these files to keep important things in context for every session. It's also useful to keep a problem solving file, where you can add problems whenever you notice them, and where the model can keep track of its progress.

\subsection{Testing \& Validation}

Testing becomes even more critical when AI generates implementation code. While AI can accelerate code generation, it cannot be trusted to ensure correctness, handle edge cases appropriately, or validate that code meets scientific standards. The rules in this section establish practices for using tests as specifications that guide AI code generation, and for leveraging AI to build and improve comprehensive test coverage. 

\airule{Implement Test-Driven Development with AI}\label{rule:tdd} Frame your test requirements as behavioral specifications before requesting implementation code, and tell the AI what success looks like through concrete test cases. This test-first approach forces you to articulate edge cases, expected inputs/outputs, and failure modes that might otherwise be overlooked \cite{beck2003test}. AI will respond better to specific test scenarios than vague functionality descriptions. By providing comprehensive test specifications, you guide the AI toward more robust, production-ready implementations. AI tools (such as chatbots or Github's Spec Kit) can help develop these specifications in a way that will optimally guide the model. Keep a close eye on the tests that are generated, since the models will often modify the tests to pass without actually solving the problem rather than generating suitable code. Be especially aware that coding agents may generate placeholder data or mock implementations that merely satisfy the test structure without validating actual logic. In many cases, the AI may insert fabricated input values or dummy functions that appear to meet acceptance criteria but do not reflect true functionality. These ``paper tests'' can be dangerously misleading, seemingly passing as tests while masking broken or incomplete logic. In addition, whenever a bug is identified during your development cycle, ask the model to generate a test that catches the bug, to ensure that it's not re-introduced in the future.

\airule{Leverage AI for Test Planning and Refinement} \label{rule:test_refine}AI is exceptionally good at identifying edge cases you might miss and suggesting comprehensive test scenarios. Feed it your function and ask it to generate tests for boundary conditions, type validation, error handling, and numerical stability. Ask it what sorts of problems your code might experience issues with, within your specified API bounds, and why those might (or might not) be relevant to address. AI can help you move beyond testing only expected behavior to robust validation that includes malformed inputs, extreme values, and unexpected conditions. Additionally, you can use AI to review your existing tests and identify gaps in coverage or scenarios you haven't considered. The AI can help you implement sophisticated testing patterns like parameterized tests, fixtures, and mocking that would be tedious to write from scratch. If you anticipate having future collaborators for your project, you may find it helpful to prioritize building testing infrastructure early. This often includes automated validation workflows, wherein you are able to test your code automatically as you integrate changes into the broader project. AI excels at generating the boilerplate for many of these sophisticated testing tools (such as GitHub Actions, pre-commit hooks, and test orchestration) that ensure your code is validated on every push.

\subsection{Code Quality \& Validation}

Even with comprehensive tests, AI-generated code requires careful human oversight to ensure correctness and appropriateness. AI models can confidently produce code that passes tests but violates domain conventions, introduces subtle bugs, or solves problems in scientifically inappropriate ways. The final rules address critical aspects of quality assurance: actively monitoring AI progress to catch problems early, and critically reviewing and continuing to iterate with all generated code to ensure it meets scientific standards.

\airule{Monitor Progress and Know When to Restart} \label{rule:supervise}It's tempting to just walk away and let the model work for a long time, but often the model will end up going down the wrong path, wasting time and tokens. You need to actively monitor what the AI is doing: Is it changing things you didn't want changed? Is it ignoring the changes you actually requested? Is it introducing new problems while trying to fix old ones? When you notice the AI heading in the wrong direction, stop it rather than letting it continue down an unproductive path. Further, sometimes the most efficient approach is recognizing when a conversation has become too convoluted with failed attempts. When this happens, review your prompt history to identify what went wrong: Were requirements unclear? Did you add conflicting constraints? Did you forget to specify critical details upfront? Starting fresh with these lessons learned often produces better results than continuing to debug within a polluted context. Clear the context and restart from externally-managed context files after adding additional details to prevent the same problem from occurring in the future. This also highlights the need for good version control; if you commit code before undertaking a major change, it's easy to simply revert to the previous commit and start over if the model goes astray. Fortunately coding agents are generally very good at writing detailed commit messages, making a commit as easy as prompting ``commit this to git’’.

\airule{Critically Review Generated Code}\label{rule:review} Be skeptical about AI's claims of success; the models tend to claim success even when they haven't really solved the problem. You always need to test the solution independently. Read and understand the code to ensure it solves problems in ways that make sense for your domain and match your prior expectation of how the problem should be solved (e.g., how you anticipated a solution looking based on your pseudocode or architecture schematics you developed in Rule \ref{rule:thinking}). AI-generated code requires careful human review to ensure scientific appropriateness, methodological soundness, and alignment with domain standards.

\airule{Refine Code Incrementally with Focused Objectives}\label{rule:increments} Once you have working, tested code, resist the temptation to ask AI to ``improve my codebase.'' Instead, approach refinement incrementally with clear, focused objectives. Be explicit about what aspect you want to improve: performance optimization, code readability, error handling, modularity, or adherence to specific design patterns. When you recognize that refinement is needed but can't articulate the specific approach (for instance, you know certain logic should be extracted into a separate function but aren't sure how), use AI to help you formulate concrete objectives before implementing changes. Describe what you are trying to achieve and ask the AI to suggest specific refactoring strategies or design patterns that would accomplish your goal, applying the same mindsets delineated in Rules \ref{rule:gather} -- \ref{rule:review} to help you along the way.

AI excels at identifying opportunities for refactoring and abstraction, such as recognizing repeated code that should be extracted into reusable functions or methods, and detecting poor programming patterns like deeply nested conditionals, overly long functions, tight coupling between components, sloppy or inconsistent variable naming conventions, and other poor patterns. When requesting refinements, specify the goal (e.g., ``extract the data validation logic into a separate function'' rather than ``make this better’’) and verify each change against your tests (while expanding your testing as you iterate to reflect updates and improvements) before moving to the next improvement. This focused approach prevents the AI from making changes that, while technically sound, don't align with your project's architectural decisions. Note that AI can inadvertently break previously working code or degrade performance while making stylistic improvements. Always test thoroughly after each incremental change, and revert if the ``improvement'' introduces problems or doesn't provide clear benefits.
\section{Discussion}

This paper presents ten rules for leveraging AI coding tools effectively in scientific computing while maintaining methodological rigor and code quality. These rules are organized around four key themes:  preparation and understanding, context engineering, testing and validation, and code quality assurance. However, we acknowledge a fundamental reality based on our experiences: even when following these rules, flawless start-to-finish interactions are the exception rather than the norm. The value of these rules lies not in guaranteeing immediate success, but in providing a framework that helps you focus on what matters most for successful interactions while also enabling you to quickly diagnose what went wrong when interactions fail, so you can iterate more effectively on your next attempt.

\paragraph{Ethical Considerations and Responsibility}

The use of AI-assisted coding raises fundamental questions about scientific accountability. When code that generates published results is partly AI-generated, who bears responsibility for errors, methodological flaws, or irreproducible outcomes? The answer must be unequivocal: the scientist. AI tools are instruments, and like any instrument in science, the researcher using them remains fully accountable for validating their outputs and ensuring methodological soundness. This responsibility cannot be delegated to the AI, regardless of how sophisticated the tool or how confident its outputs appear. Researchers must ensure their AI-assisted code is reproducible, well-documented, and scientifically appropriate. When AI generates code that implements a statistical method or analytical pipeline, the researcher must understand that implementation well enough to defend its appropriateness, explain its limitations, and troubleshoot unexpected results. ``AI wrote it’’ is not a valid defense for flawed methodology or incorrect results. Transparency about AI usage in methods sections, while important, does not diminish this responsibility.

Beyond individual accountability, broader ethical concerns demand serious consideration. The environmental costs of training and running large language models are substantial and measurable \cite{faiz2024llmcarbon,ren2024environmental}. These systems consume enormous amounts of energy and computational resources, raising questions about the sustainability of widespread AI adoption. Further, intellectual property questions surrounding AI training on open-source code and the ownership of AI-generated code remain legally and ethically unsettled \cite{nordemann2022copyright,buick2025copyright,blaszczyk2024ai,kretschmer2024copyright}. Courts have yet to definitively rule on whether training on copyrighted code constitutes fair use, whether AI-generated code can be copyrighted, and who owns the rights to such code when models have been trained on proprietary or licensed material. These are fundamental ethical and legal challenges that the scientific community must grapple with as AI tools become embedded in research infrastructure. While these complex issues merit a dedicated treatment beyond our scope here, researchers should recognize that using AI coding tools involves participating in systems with significant unresolved ethical dimensions.

\paragraph{Guardrails for Autonomous Agents}

Autonomous coding agents can make extensive changes across a codebase with minimal human intervention, dramatically accelerating development but introducing risks if not properly constrained. The primary danger lies in granting agents too much control without appropriate safeguards. An agent given broad permissions might break existing functionality, introduce security vulnerabilities, or violate architectural principles while reporting success.

We recommend several guardrails. First, use containerized or sandboxed environments for agent-driven development, isolating agent operations from production systems \cite{wiebels2021leveraging}. Second, commit working code before allowing agent changes, enabling easy rollback. Third, learn how to properly configure agents with explicit constraints about what they can modify and what actions require human approval. Fourth, maintain active monitoring rather than allowing unsupervised operation, as discussed in Rule \ref{rule:supervise}. For individual projects, consider project-specific containers where each agent operates in an isolated environment with restricted file access. As autonomous agents become more capable, developing clear and safe practices for constraining and monitoring their behavior will become increasingly important for maintaining scientific rigor and system safety.

\paragraph{Limitations and Future Directions}

We acknowledge that we are operating in a rapidly evolving technological landscape. For reference, GPT-3 (2020) had a context window of 2,048 tokens \cite{brown2020language}, GPT-4 (2023) expanded this to variants with tens of thousands of tokens \cite{openai2023gpt4}, and current state-of-the-art models like Gemini 2.5 Pro (2025) \cite{google2024gemini2} can operate with context windows of millions of tokens. In light of this rapid evolution, we have intentionally focused on principles and practices that remain relevant across different AI capabilities. We believe our proposed rules emphasize fundamental skills (domain knowledge, problem decomposition, critical review) and strategies (context management, test-driven development, incremental refinement) that apply regardless of specific tools, and thus far have proven useful throughout the evolution of AI models to-date. We have deliberately avoided prescriptive recommendations and strategies tied to specific models, as these would quickly become outdated. Future advances may change which practices prove most valuable, but we believe these rules provide a useful framework for current practice that will remain adaptable as technology matures.

We also anticipate substantial evolution in how scientists acknowledge the role of AI in their work. As AI coding becomes standard practice, we expect clearer community expectations for documenting AI tool usage and validating AI-generated code; this may include citations of specific systems, disclosure of prompting approaches, detailed validation procedures in methods sections, and heightened expectations regarding testing, validation, and reproducibility of code derivatives. The practices we recommend (systematic context building, comprehensive testing, and critical validation) may provide a foundation for informing and meeting these emerging accountability standards.

\paragraph{Further Reading}
The rules presented in this paper provide a framework for using AI tools effectively in scientific computing, but they build upon foundational concepts in software development, reproducibility, and design. The following resources introduce key concepts and practices that support the development of programming skills necessary for effective AI-assisted coding, and may help readers deepen their understanding of principles underlying the rules.
\begin{enumerate}[leftmargin=*]
    \item LeVeque, Randall J., et al. Reproducible Research for Scientific Computing: Tools and Strategies. \textit{Computing in Science \& Engineering}, vol. 14, no. 4, 2012, pp. 13-17. \cite{leveque2012reproducible} This article establishes best practices for reproducible scientific computing that inform our emphasis on documentation, testing, and validation throughout the rules.
    \item Ousterhout, John. \textit{A Philosophy of Software Design}. 2nd ed., Yaknyam Press, 2021. \cite{ousterhout2021philosophy} This book articulates core principles of software design, including abstraction and modularity, that inform our guidance in Rules \ref{rule:thinking} and \ref{rule:context} on specifying context for scientific problems and thinking through incremental refinement in Rule \ref{rule:increments}.
    \item Poldrack, Russell A. \textit{Better Code, Better Science}. \href{https://poldrack.github.io/BetterCodeBetterScience/frontmatter.html}{poldrack.github.io/BetterCodeBetterScience}. Accessed Sept 10, 2025. \cite{poldrack2024better} This comprehensive guide introduces AI tools in scientific workflows and provides practical guidance that complements the principles outlined in our rules.
    \item Felleisen, Matthias, et al. \textit{How to Design Programs: An Introduction to Programming and Computing}. 2nd ed., MIT Press, 2018, \href{https://htdp.org}{htdp.org}. \cite{felleisen2018design} This text emphasizes systematic problem decomposition and design principles that underpin the distinction between problem framing and coding discussed in Rule \ref{rule:framing_vs_coding}.
    \item Beck, Kent. \textit{Test-Driven Development: By Example}. Addison-Wesley, 2003. \cite{beck2003test} This book provides the foundational methodology for test-driven development discussed in Rule \ref{rule:tdd}, demonstrating how writing tests before implementation can help you to develop more robust and maintainable code.
    \item Wiebels, Kristina, and David Moreau. Leveraging Containers for Reproducible Psychological Research. \textit{Advances in Methods and Practices in Psychological Science}, vol. 4, no. 2, 2021. \cite{wiebels2021leveraging} This paper demonstrates how containerization supports reproducible research, directly relevant to the guardrails for autonomous agents discussed in our Discussion section.
\end{enumerate}

\paragraph{Acknowledgements} An initial framework of 20 rules and focus points was developed by EWB and RAP (10 rules each). These rules were streamlined into 10 rules and bullet points with assistance from Claude (Anthropic) \cite{anthropic2025claude45}, which were then authored and iteratively refined by the research team into the content and recommendations presented herein. This work was supported by a grant from the Sloan Foundation to RAP (G-2025-25270).

\paragraph{Code and Data Availability Statement} The interactive Jupyter Book referenced in this manuscript is available at \href{https://poldracklab.org/10sr_ai_assisted_coding}{poldracklab.org/10sr\_ai\_assisted\_coding} and permanently indexed at \cite{bridgeford2025aiassist}.

\bibliographystyle{unsrtnat}
\bibliography{article_bib}

\begin{thebibliography}{25}
\providecommand{\natexlab}[1]{#1}
\providecommand{\url}[1]{\texttt{#1}}
\expandafter\ifx\csname urlstyle\endcsname\relax
  \providecommand{\doi}[1]{doi: #1}\else
  \providecommand{\doi}{doi: \begingroup \urlstyle{rm}\Url}\fi

\bibitem[Chen et~al.(2021)Chen, Tworek, Jun, Yuan, Pinto, Kaplan, Edwards, Burda, Joseph, Brockman, et~al.]{chen2021evaluating}
Mark Chen, Jerry Tworek, Heewoo Jun, Qiming Yuan, Henrique Pond{\'e} de~Oliveira Pinto, Jared Kaplan, Harri Edwards, Yura Burda, Nicholas Joseph, Greg Brockman, et~al.
\newblock Evaluating large language models trained on code.
\newblock \emph{arXiv preprint arXiv:2107.03374}, 2021.

\bibitem[Peng et~al.(2023)Peng, Kalliamvakou, Cihon, and Demirer]{peng2023impact}
Sida Peng, Eirini Kalliamvakou, Peter Cihon, and Mert Demirer.
\newblock The impact of ai on developer productivity: Evidence from github copilot.
\newblock \emph{arXiv preprint arXiv:2302.06590}, 2023.

\bibitem[Kalliamvakou et~al.(2024)Kalliamvakou, Ziegler, Li, Rice, Rifkin, Simister, Sittampalam, and Aftandilian]{kalliamvakou2024measuring}
Eirini Kalliamvakou, Albert Ziegler, X~Alice Li, Andrew Rice, Devon Rifkin, Shawn Simister, Ganesh Sittampalam, and Edward Aftandilian.
\newblock Measuring github copilot's impact on productivity.
\newblock \emph{Communications of the ACM}, 67\penalty0 (3):\penalty0 54--63, 2024.

\bibitem[Becker and Rush(2025)]{becker2025measuring}
Joel Becker and Nate Rush.
\newblock Measuring the impact of early-2025 ai on experienced open-source developer productivity.
\newblock \emph{arXiv preprint arXiv:2507.09089}, 2025.

\bibitem[Harding et~al.(2024)Harding, Kloster, and {GitClear}]{gitclear2024}
Bill Harding, Matthew Kloster, and {GitClear}.
\newblock Ai copilot code quality: 2023 data suggests downward pressure on code quality.
\newblock GitClear Research Report, 2024.
\newblock URL \url{https://gwern.net/doc/ai/nn/transformer/gpt/codex/2024-harding.pdf}.

\bibitem[Poldrack(2024)]{poldrack2024better}
Russell~A. Poldrack.
\newblock Better code, better science.
\newblock \url{https://poldrack.github.io/BetterCodeBetterScience/}, 2024.
\newblock Accessed: 2025-09-10. doi: 10.5281/zenodo.17407478.

\bibitem[{Chroma}(2024)]{chroma2024context}
{Chroma}.
\newblock Context rot: How increasing input tokens impacts llm performance.
\newblock Technical report, Chroma, 2024.
\newblock URL \url{https://research.trychroma.com/context-rot}.

\bibitem[Bridgeford(2025)]{bridgeford2025aiassist}
Eric Bridgeford.
\newblock Poldracklab/10sr\_ai\_assisted\_coding: V0.0, oct 2025.
\newblock URL \url{https://doi.org/10.5281/zenodo.17398109}.

\bibitem[Martin(2008)]{martin2008clean}
Robert~C. Martin.
\newblock \emph{Clean Code: A Handbook of Agile Software Craftsmanship}.
\newblock Prentice Hall, Upper Saddle River, NJ, 2008.
\newblock ISBN 9780132350884.

\bibitem[Beck(2003)]{beck2003test}
Kent Beck.
\newblock \emph{Test-Driven Development: By Example}.
\newblock Addison-Wesley, Boston, MA, 2003.

\bibitem[Faiz et~al.(2024)Faiz, Kaneda, Wang, Osi, Sharma, Chen, and Jiang]{faiz2024llmcarbon}
Ahmad Faiz, Sotaro Kaneda, Ruhan Wang, Rita~Chukwunyere Osi, Prateek Sharma, Fan Chen, and Lei Jiang.
\newblock Llmcarbon: Modeling the end-to-end carbon footprint of large language models.
\newblock In \emph{The Twelfth International Conference on Learning Representations (ICLR)}, 2024.

\bibitem[Ren et~al.(2024)Ren, Tomlinson, Black, and Torrance]{ren2024environmental}
Shaolei Ren, Bill Tomlinson, Rebecca~W Black, and Andrew~W Torrance.
\newblock Reconciling the contrasting narratives on the environmental impact of large language models.
\newblock \emph{Scientific Reports}, 14\penalty0 (1):\penalty0 26310, 2024.
\newblock \doi{10.1038/s41598-024-76682-6}.

\bibitem[Nordemann and Pukas(2022)]{nordemann2022copyright}
Jan~Bernd Nordemann and Jonathan Pukas.
\newblock Copyright exceptions for {AI} training data---will there be an international level playing field?
\newblock \emph{Journal of Intellectual Property Law \& Practice}, 17\penalty0 (12):\penalty0 973--974, 2022.
\newblock \doi{10.1093/jiplp/jpac106}.

\bibitem[Buick(2025)]{buick2025copyright}
Adam Buick.
\newblock Copyright and {AI} training data---transparency to the rescue?
\newblock \emph{Journal of Intellectual Property Law \& Practice}, 20\penalty0 (3):\penalty0 182--192, 2025.
\newblock \doi{10.1093/jiplp/jpae102}.

\bibitem[Blaszczyk et~al.(2024)Blaszczyk, McGovern, and Stanley]{blaszczyk2024ai}
Matt Blaszczyk, Geoffrey McGovern, and Karlyn~D. Stanley.
\newblock Artificial intelligence impacts on copyright law.
\newblock Perspective PEA3243-1, RAND Corporation, Santa Monica, CA, 2024.

\bibitem[Kretschmer et~al.(2024)Kretschmer, Margoni, and Oru\c{c}]{kretschmer2024copyright}
Martin Kretschmer, Thomas Margoni, and Pinar Oru\c{c}.
\newblock Copyright law and the lifecycle of machine learning models.
\newblock \emph{{IIC} - International Review of Intellectual Property and Competition Law}, 55:\penalty0 110--138, 2024.
\newblock \doi{10.1007/s40319-023-01419-3}.

\bibitem[Wiebels and Moreau(2021)]{wiebels2021leveraging}
Kristina Wiebels and David Moreau.
\newblock Leveraging containers for reproducible psychological research.
\newblock \emph{Advances in Methods and Practices in Psychological Science}, 4\penalty0 (2), 2021.
\newblock \doi{10.1177/25152459211017853}.

\bibitem[Brown et~al.(2020)Brown, Mann, Ryder, Subbiah, Kaplan, Dhariwal, Neelakantan, Shyam, Sastry, Askell, et~al.]{brown2020language}
Tom~B. Brown, Benjamin Mann, Nick Ryder, Melanie Subbiah, Jared Kaplan, Prafulla Dhariwal, Arvind Neelakantan, Pranav Shyam, Girish Sastry, Amanda Askell, et~al.
\newblock Language models are few-shot learners.
\newblock In \emph{Advances in Neural Information Processing Systems}, volume~33, pages 1877--1901, 2020.

\bibitem[{OpenAI} et~al.(2023)]{openai2023gpt4}
{OpenAI} et~al.
\newblock {GPT-4} technical report.
\newblock Technical report, OpenAI, 2023.
\newblock arXiv:2303.08774.

\bibitem[{Google DeepMind}(2024)]{google2024gemini2}
{Google DeepMind}.
\newblock Introducing {Gemini} 2.0: Our new {AI} model for the agentic era.
\newblock Google Blog, December 2024.
\newblock URL \url{https://blog.google/technology/google-deepmind/google-gemini-ai-update-december-2024/}.

\bibitem[LeVeque et~al.(2012)LeVeque, Mitchell, and Stodden]{leveque2012reproducible}
Randall~J LeVeque, Ian~M Mitchell, and Victoria Stodden.
\newblock Reproducible research for scientific computing: Tools and strategies for changing the culture.
\newblock \emph{Computing in Science \& Engineering}, 14\penalty0 (4):\penalty0 13--17, 2012.
\newblock \doi{10.1109/MCSE.2012.38}.

\bibitem[Ousterhout(2021)]{ousterhout2021philosophy}
John Ousterhout.
\newblock \emph{A Philosophy of Software Design}.
\newblock Yaknyam Press, 2 edition, 2021.

\bibitem[Felleisen et~al.(2018)Felleisen, Findler, Flatt, and Krishnamurthi]{felleisen2018design}
Matthias Felleisen, Robert~Bruce Findler, Matthew Flatt, and Shriram Krishnamurthi.
\newblock \emph{How to Design Programs: An Introduction to Programming and Computing}.
\newblock MIT Press, 2 edition, 2018.
\newblock URL \url{https://htdp.org}.

\bibitem[{Anthropic}(2025)]{anthropic2025claude45}
{Anthropic}.
\newblock Claude sonnet 4.5.
\newblock Anthropic Product Release, 2025.
\newblock URL \url{https://www.anthropic.com/claude/sonnet}.

\bibitem[Vaswani et~al.(2017)Vaswani, Shazeer, Parmar, Uszkoreit, Jones, Gomez, Kaiser, and Polosukhin]{vaswani2017attention}
Ashish Vaswani, Noam Shazeer, Niki Parmar, Jakob Uszkoreit, Llion Jones, Aidan~N Gomez, {\L}ukasz Kaiser, and Illia Polosukhin.
\newblock Attention is all you need.
\newblock In \emph{Advances in neural information processing systems}, volume~30, 2017.

\end{thebibliography}

\newpage

\appendix

\section{Background concepts and vocabulary}
\label{app:background}

To navigate the challenges of AI-assisted coding effectively, researchers should be familiar with several key concepts that underpin these tools:

\begin{itemize}[leftmargin=*]
\item \textbf{Large Language Models (LLMs)} are neural networks trained on vast text corpora that generate text by predicting sequences of tokens, basic units of text processing that typically represent words, parts of words, or individual characters \cite{vaswani2017attention}. For instance, the word ``unhappily'' might be tokenized as ``un'', ``\#\#happi'', ``\#\#ly'', where \#\# marks tokens that are not the start of a word.

\item \textbf{Context windows} define the maximum number of tokens an LLM can consider when generating responses. State-of-the-art models typically handle hundreds of thousands to millions of tokens, constraining how much code and documentation they can simultaneously process. When context limits are exceeded, models lose track of earlier information. Even when information is contained within the context window, attention to mid-document details can degrade (``lost in the middle''), especially for models with very large context windows; this phenomenon is known as \textbf{context rot}. For an example of context rot, see \cite{chroma2024context}.

\item \textbf{In-context learning} allows models to adapt their behavior based on examples and instructions provided within the current conversation, without permanent changes to the underlying model. This enables direction of model behavior through strategic provision of examples and formatting of instructions.

\item \textbf{Prompting} encompasses techniques for structuring inputs to elicit desired outputs, including clear requirement specification, strategic provision of examples, and structured formatting. Effective prompting can dramatically improve code quality and relevance.

\item \textbf{Test-driven development} involves writing tests before implementation to specify expected behavior and validate correctness, a practice that becomes even more critical when AI generates the implementation code. Test driven development is detailed in \cite{beck2003test}.
\end{itemize}

\subsection{Sharing context}

AI coding tools range from conversational interfaces like ChatGPT to interactive assistants like GitHub Copilot to autonomous coding agents like Cursor. Each presents unique challenges for maintaining project context across sessions. Most AI systems are stateless, meaning they forget previous interactions, while others have limited understanding of broader project requirements. This creates two critical problems: context fragmentation, where important project details are lost between sessions, and iteration drift, where AI assistance gradually diverges from intended goals without proper oversight.

\textbf{Externally-managed context files} help address these limitations by providing persistent information across AI interactions. These include:

\begin{itemize}[leftmargin=*]
\item \textbf{Memory files} contain project-specific information like architectural decisions, software development standards and practices, and lessons learned that persist between interactions. They prevent repetition of past mistakes and ensure each new AI session starts with relevant context.

\item \textbf{Constitution files} establish non-negotiable principles governing AI behavior throughout development, such as security requirements or methodological constraints.
\end{itemize}

Together, these tools can help transform AI interactions into consistent, goal-directed collaboration by providing the persistent context and boundaries that AI systems lack natively.

\end{document}